\documentclass[useAMS,usenatbib]{mnras}
\usepackage{amsmath}
\usepackage{amssymb}
\usepackage{graphicx}
\usepackage{multirow}
\usepackage{color}
\usepackage{array}
\usepackage{graphicx}
\title[Geometrical optic approach for AGN corona]{Effects of the refractive index of the X-ray corona on the emission lines in AGN}

\author[P. Chainakun et al.]{P. Chainakun$^{1,2}$\thanks{E-mail: \href{mailto:pchainakun@g.sut.ac.th}{pchainakun@g.sut.ac.th}}, A. Watcharangkool$^3$, A. J. Young$^4$  \\
$^1$School of Physics, Institute of Science, Suranaree University of Technology, Nakhon Ratchasima 30000, Thailand\\
$^2$Centre of Excellence in High Energy Physics and Astrophysics, Suranaree University of Technology, Nakhon Ratchasima 30000, Thailand\\
$^3$National Astronomical Research Institute of Thailand, Chiang Mai 50200, Thailand\\
$^4$H. H. Wills Physics Laboratory, Tyndall Avenue, Bristol BS8 1TL, UK}

\date{Accepted XXX. Received YYY; in original form ZZZ}

\pubyear{xxxx}

\begin{document}
%\date{}

\pagerange{\pageref{firstpage}--\pageref{lastpage}} \pubyear{2015}

\maketitle

\label{firstpage}

\begin{abstract}
X-ray reflection from an accretion disc produces characteristic emission lines allowing us to probe the innermost regions in AGN. We investigate these emission lines under a framework of Riemannian geometrical optics where the corona has a refractive index of $n \neq 1$. The empty space outside is a vacuum with $n=1$. The Kerr metric is modified to trace the light rays that are bent due to not only the gravity of the black hole, but also the effects of coronal plasma dependent on $n$. The choice of $n$ alters the null geodesics, producing the effect which is analogous to the light deflection. For the corona with $n>1$, the disc on the far side within the corona covers a larger area on the observer’s sky, enhancing the blue wing of the line and producing more flux difference between the blue peak and extended red tail. The inverse effects are seen when $n<1$. Moreover, the corona with $n>1$ and $n<1$ could induce extra shifts in the blue wing ($\Delta g_{\rm max}$) to higher and lower energy, respectively. These effects are more prominent when the inclination angle is $\gtrsim 60^{\circ}$ and the corona extends to $\gtrsim 5r_{\rm g}$. To obtain the deviation of the line shift of $\Delta g_{\rm max} \gtrsim 0.01$, the difference between the refractive index of the corona and that of the empty space must be $\Delta n \gtrsim 0.5\%$. Finally, the lensing corona can influence the arrival time of photons that may affect the observed variability of these emission lines.

\end{abstract}


\section{Introduction}

The X-ray continuum in AGN is produced when the optical and UV photons from an accretion disc are Compton up-scattered by relativistic hot electrons inside the corona. The X-ray irradiation back onto the disc surface can be reprocessed and scattered off the disc producing the reflection spectrum where the characteristic features such as photoelectric absorption and fluorescent lines are imprinted on \citep{Fabian2000, Reynolds2003}. Due to the high fluorescent yield and large cosmic abundance, the strongest observable feature among the X-ray fluorescence lines from neutral material is the Fe K$\alpha$ line at $\sim 6.4$~keV. This Fe K$\alpha$ emission was first discovered by \cite{Tanaka1995} in MCG-6-30-15, before being observed in many other AGN \citep[e.g.][and references therein]{Nandra2007}. 

The observed spectral lines are naturally broadened and distorted, suggesting that they are produced at the innermost region of the accretion disc where the relativistic effects are very strong \citep[e.g.][]{Fabian2000}. The line width is broadened due to the Keplerian rotation of the disc and the broadest lines are produced closest to the central black hole where the materials move fastest. The emission from the approaching and receding sides of the disc with respect to an observer contributes to the blueshifted and redshifted peaks of the lines, whose amplitude can be further enhanced and suppressed, respectively, due to the transverse Doppler shift and Doppler beaming. Moreover, due to the general relativistic effects, the emission photons are gravitationally redshifted since they lose their energy when escaping from the potential well of the black hole. Ultimately, the line shapes become a broad and skewed profile with a narrow blue peak and wide faint red tail.

The studies of these emission lines as a diagnostic tool for probing the central regions of the compact system have been carried out extensively \citep[see][for a review]{Reynolds2003}. Under the weak-ﬁeld approximation, the analytical approach to study the Fe K$\alpha$ lines is possible \citep[e.g.][]{Chen1989}. The realistic line profiles, however, can be computed only numerically by integrating photon paths along the geodesics around the black hole \citep{Fabian1989, Laor1991, Karas1992, Bromley1997, Fanton1997}. The characteristic features of the lines strongly depend on the ionization state of the disc that determines how the incident X-ray continuum is reprocessed \citep{George1991, Matt1993, Ross1999, Fabian2000}. Nowadays, there are several public reflection models for simulating the full reflection spectrum, not only the Fe K$\alpha$ line, from an ionized slab of gas irradiated by an external X-ray source \citep[e.g.][]{Ross2005, Garcia2013, Garcia2014}. 

While the broad spectral line has an inner-disc reflection origin, the nature of the corona itself is less clear. Based on the reverberation lag analysis which measures the time delays associating with the additional light crossing distance travelled by the reflection photons \citep[see][for a review]{Uttley2014, Cackett2021}, the AGN corona, with assumed lamp-post configuration, is found to be close to the central black hole \citep[e.g.][]{Wilkins2013, Cackett2014, Emmanoulopoulos2014, Chainakun2016, Epitropakis2016, Ingram2019, Caballero2020}. Extended corona models have also been developed \citep[e.g.][]{Wilkins2016, Chainakun2017, Chainakun2019b}. An interplay between the geometry of the disc and the corona determines the emissivity profile of the disc which, in turn, affects the shape of the emission lines \citep[e.g.][]{Jovanovi2010, Wilkins2012, Jovanovi2016}. 

Here, we investigate the properties of the line profiles produced under the framework of geometrical optics that explains the photon trajectories using the ray-like properties of light. The corona is assumed to have a spherical shape, as in \cite{Chainakun2019b}, that requires only the radial size to constrain its geometry. An empty space (i.e. a vacuum) outside the corona is a medium with the refractive index $n=\sqrt{\varepsilon_0\mu_0}=1$. On the other hand, the corona is treated as an optical medium with permittivity $\varepsilon$ and permeability $\mu$,  which could yield either $n>1$ or $n<1$. Therefore, the refraction of light rays occurs at the coronal shell due to the change of permittivity and permeability of the medium. The use of geometrical optical approximations in the context of ray tracing and in the presence of dielectric has been introduced by \cite{Gordon1923}, where the optical metric including the refractive index was derived. Here, we follow the original work of \cite{Gordon1923} by modifying the Kerr metric so that the light deflection can be understood in terms of the effect of $n$ that alters the null geodesics in the modified Kerr spacetime. The observed line profiles are computed using the ray-tracing technique that numerically follows the light trajectories along the modified geodesics around a spinning black hole. Treating the AGN corona as a geometrical optic object introduces additional light-bending effects that could manifest the emission-line profiles as seen by a distant observer.

Previous studies on the influence of the medium on the propagation of light rays around compact objects are mainly contributed in the field of gravitational lensing with a presence of plasma \citep[e.g.,][]{Bisnovatyi2010, Morozova2013, Er2014, Rogers2017, Perlick2017, Tsupko2020, Tsupko2021}. It was suggested that general relativistic and plasma effects in the vicinity of a compact object could generate an observable polarization of photons travelling through a plasma \citep{Broderick2003, Broderick2004}. \cite{Babar2020} modified the Kerr-Newman metric using the Synge formalism and the Hamilton-Jacobi method, by embedding the refractive index $n$ directly into the Hamilton-Jacobi equation, and traced the influence of plasma to the generic photon trajectories from the corresponding equations of motion. 

We note that different authors may employ different modifications and applications in this context, but it is general to apply the refractive index to the metric either in the flat or the curved spacetime \citep[e.g.][]{Gordon1923, Zhu1997}. For example, \cite{Babar2020} derived the Hamilton-Jacobi equation and equations of motion of photons in the case of a Kerr-Newman black hole surrounded by a plasma. Their approach can be traced back to the presence of $n$ in the metric too. In fact, this work is comparable to that of, e.g., \cite{Zhu1997}, \cite{Chen2009} and \cite{Babar2020} in the way that $n$ is directly implemented into the metric, hence into the equations of motion. While \cite{Babar2020} traced the photon orbits in the vicinity of black holes in the presence of plasma, we trace the photons directly backward from the observer’s sky to the disc. We focus on the X-ray coronal effects that may induce the light deflection and cause the deviation of the observed emission lines in AGN. How the line profiles change with the refractive index of the corona is investigated. 

The modified Kerr metric in the context of Riemannian geometrical optics and related assumptions are presented in Section 2. Equations of motion in modified geodesics and the calculations of the observed emission lines are presented in Section 3. The results of line profiles dependent on a variety of coronal radii, inclinations of the system, and indices of the disc emissivity are presented in Section 4, following by the discussion in Section 5. The conclusion is given in Section 6. Detailed calculations of the modification of the metric used here are presented in Appendix A.

\section{Riemannian Geometrical Optics and modified Kerr metric}

\begin{figure}
    \centerline{
        \includegraphics[width=0.5\textwidth]{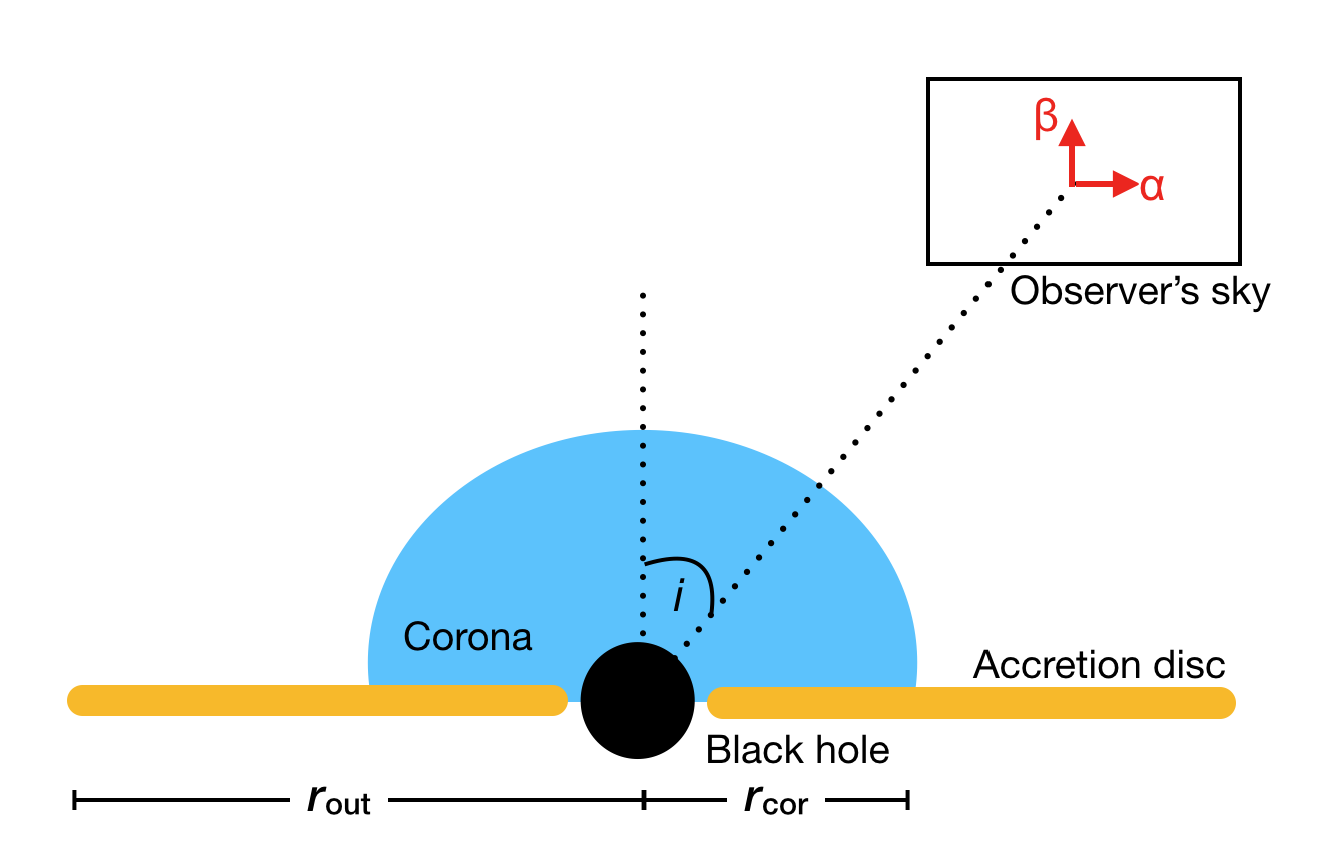}
    }
    \caption{Schematic of the AGN corona model and the setup of our geometry.
    \label{geometry1}}
\end{figure}

In geometrical optics, reflection and refraction of light follow Fermat’s principle: a light ray always takes a path that requires the least amount of time \citep[e.g.][]{Giannoni2002}. For simplicity, we treat the X-ray corona as a homogenous non-dispersive medium whose optical properties are determined by the refractive index of the corona. The corona has an infinitesimally small shell causing time delays which depend on the entry point of the light ray passing through the coronal shell. The accretion disc is geometrically thin and optically thick \citep{Shakura1973} where the innermost stable circular orbit (ISCO) is defined by the black hole spin, $a$. We fixed $a=0.998$ so $r_{\text{ISCO}}\sim 1.23r_{\text g}$, hence the region inside $r_{\text{ISCO}}$ is very small and is negligible. Note that $r_{\text{g}} = GM/c^2$ where $G$ is the gravitational constant, $M$ is the black hole mass and $c$ is the speed of light. Our geometric set-up is presented in Fig.~\ref{geometry1}. The radius of the spherical corona is given by $r_{\rm cor}$. Observed flux affected by the lensing corona is mostly from the inner regions, so to reduce the computational time, the outer radius of the disc is fixed at $r_{\rm out} = 50r_{\rm g}$. We create a distant photographic plate representing an observer's sky spanned by the Cartesian coordinates ($\alpha$, $\beta$). The origin of the observer's sky is fixed at ($\alpha=0$, $\beta=0$) while the parameter $i$ is the inclination of the system.

In this scenario, the light rays travel in an empty space except inside the corona. Assuming that the refractive index $n(r)$ is a function of radius; 
\begin{equation}
n(r) =
\begin{cases}
\sqrt{\epsilon\mu} & \text{;} \; \; r \leq r_{\rm cor} \\ 
1 & \text{;} \; \; r > r_{\rm cor}  
\label{rf-index}
\end{cases}
\end{equation}
where $\varepsilon$ and $\mu$ are permittivity and permeability of the corona. Note that natural units $G=c=1$ are used. Outside the corona ($r  > r_{\rm cor}$), the refractive index $n=1$. Inside the corona ($r  \leq r_{\rm cor}$), the refractive index can be $n>1$ or $n<1$. In fact, $n$ can be a complex number that contains the real and imaginary parts explaining the deflection and absorption, respectively. The real and imaginary parts then produce independent effects that can be considered independently. Furthermore, for relativistic electron gas, the refractive index can even be a negative value \citep{deCarvalho2016}. One can choose different assumptions for varying refractive index but, for simplicity, we first consider only the real part of $n$, as in e.g. \cite{Babar2020}, and assume it to be constant throughout the interior of the corona. In other words, we focus only on the effects of coronal plasma on the light deflection at the coronal boundary without absorption. When the light enters the corona, it experiences the geometry difference from those of curved spacetime in vacuum as their refractive indices are dissimilar. 

In a homogenous medium in Kerr spacetime, the light rays propagate along the Kerr geodesics and bend at the interface between two different media where the refractive index changes. We employ the optical metric motivated by \cite{Gordon1923} to describe the light rays travelling outside and inside the corona 
\begin{equation}
    \begin{split}
        {\rm d}s^{2} = & -\bigg{(}1-\frac{2Mr}{\Sigma}\bigg{)}n(r)^{-2}{\rm d}t^{2} - \bigg{(}\frac{4Mar\sin^{2}\theta}{n(r)\Sigma}\bigg{)}{\rm d}t{\rm d}\varphi \\
         & + \bigg{(}\frac{\Sigma}{\Delta}\bigg{)}{\rm d}r^{2} + \Sigma{\rm d}\theta^{2} + \bigg{(}\frac{\Delta_\theta\sin^{2}\theta}{\Sigma}\bigg{)}{\rm d}\varphi^{2}~,   
    \end{split}
    \label{Kerr}
\end{equation}
for which we defined
\begin{align}
\Sigma & = r^2+a^2cos^{2}\theta~, \\
\Delta & = r^2+a^2-2Mr~, \\
\Delta_\theta & = (r^{2}+a^{2})^{2} -a^{2}\Delta\sin^{2}\theta~,
\end{align}
where $n(r)$ is the refractive index that changes accordingly to the conditions set in eq.~\ref{rf-index}. When $n(r)=1$, the metric in eq.~\ref{Kerr} reduces to the Kerr metric that explains the trajectories of light rays in the vacuum space. The angular momentum of the black hole points along the rotational axis towards $\theta=0$, while $\varphi$ is the azimuth angle measured in the accretion disc plane. 

The factor $n(r)$ in the ${\rm d}t$ term yields a non-homogenous expanding/contracting spacetime. This application is analogically similar to the Friedmann-Robertson-Walker (FRW) metric that explains a homogeneous isotropic expanding spacetime which can cause superluminal expansion but does not violate relativity. This modification results in the metric that is similar to, e.g., \cite{Gordon1923} and \cite{Chen2009} in the case of co-moving fluid in Minkowski, Schwarzshild, and FRW spacetime. In this context, different $n$ leads to different obtained null geodesics that alter the light trajectories, producing the comparable effects of light deflection at the coronal boundary. More detailed calculations and justification of the modified Kerr metric is given in Appendix A.

\section{Light rays propagating in modified geodesics}

\subsection{Equations of motion}

Here, we derive the equations of motion of photons along the modified Kerr geodesics presented in eq.~\ref{Kerr}. We measure distance and time in gravitational units of $r_{\text{g}} = GM/c^2$ and $t_{\text{g}} = GM/c^3$, respectively. To obtain equations of motion, we consider constants of motion which are the total energy, $E$, and the angular momentum, $L$: 
\begin{align}
E=& A\dot{t}+B\dot{\phi}~, \\
L=& -B\dot{t}+C\dot{\phi}~,
\end{align}
where $A=\left( 1-\frac{2Mr}{\Sigma}\right)n^{-2},  B=2Mar\sin^2{\theta}/n\Sigma$, and $ C=\Delta_\theta\sin^2{\theta}/\Sigma$. Note that $AC+B^2=\sin^2{\theta}\Delta/n^2$ and the equations of motion for coordinate $t$ and $\varphi$ read
\begin{align}
\Sigma \frac{dt}{d\lambda} & =n^2\left[ -a\left(a\sin^2{\theta}E - \frac{L}{n}\right) + \frac{(r^2+a^2)}{\Delta}T_n\right]~, \label{mot1}\\
\Sigma \frac{d\varphi}{d\lambda} & =n\left[ -aE + \frac{L}{\sin^2{\theta}} + \frac{a}{\Delta}T_n\right]~,
\label{mot2}
\end{align}
where $T_n  = E(r^2+a^2)-aL/n$. For the case of a massless particle, the parameter $\lambda$ is the proper time. To find the rest equations of motion, we use the Hamilton-Jacobi equation with an ansatz
\begin{equation}
S=-Et+L\phi+S^{(r)}+S^{(\theta)}~,
\end{equation}
which yields
\begin{align}
0=& -E\dot{t}+L\dot{\phi}+\left(\frac{\partial S^{(r)}}{\partial r}\right)^2+\left(\frac{\partial S^{(\theta)}}{\partial \theta}\right)^2 \nonumber \\
 =&~\left(\Delta\left(\frac{\partial S^{(r)}}{\partial r}\right)^2-2naLE - \frac{(r^2+a^2)En^2}{\Delta}T_n+\frac{anL}{\Delta}T_n \right) \nonumber \\
 &~+ \left(\left(\frac{\partial S^{(\theta)}}{\partial \theta}\right)^2 + \frac{L}{\sin^2{\theta}}\right)+a^2n^2\sin^2{\theta}E^2~.
\end{align}
Notice that the expression inside the first bracket is a function of $r$, hence one may add $-a^2n^2E^2-L^2$ and define a function
\begin{align}
Q(r)=&~ -\Delta\left(\frac{\partial S^{(r)}}{\partial r}\right)^2 -(anE-L)^2 \nonumber \\
&~+ \left(\frac{(r^2+a^2)En^2}{\Delta}-\frac{anL}{\Delta}\right)T_n \nonumber \\
       =&~ \left(\frac{\partial S^{(\theta)}}{\partial \theta}\right)^2 + L^2\cot{\theta}^2-a^2n^2E^2\cos^2{\theta}~.
\end{align}
Since $S$ is the solution of the Hamilton-Jacobi equation of motion, we have 
\begin{align}
\frac{\partial S^{(\theta)}}{\partial \theta}=&~p_\theta=\Sigma\dot{\theta}~, \\
\frac{\partial S^{(r)}}{\partial r}=&~p_r=\frac{\Sigma}{\Delta}\dot{r}~,
\end{align}
hence 
\begin{equation}
Q(r)=Q_C+a^2E^2\cos^2\theta\left(1-n^2\right)~,
\label{Qr}
\end{equation}
where $Q_C$ denotes the usual Carter constant. Then, we can derive the equations of motion for the coordinate $r$ and $\theta$ as follows,  
\begin{align}
\Sigma\frac{dr}{d\lambda}=&~\pm\sqrt{n^2T_n^2-\Delta\left[Q(r)+(L-anE)^2\right]}~, \label{mot3} \\
\Sigma\frac{d\theta}{d\lambda}=&~\pm\sqrt{Q_C+\cos^2{\theta}\left(a^2E^2-\frac{L^2}{\sin^2{\theta}}\right)}~. \label{mot4}
\end{align}

Although in the context of waveguide, $n < 1$ could imply the speed larger than $c$. This is the wavefront speed and $n < 1$ is likely possible, and probably a commonplace, for the plasma. Note that the equations of motion are derived from the modified Kerr metric through the consideration of $ \mathcal{L} =g_{\alpha \beta}\dot{x}^{\alpha}\dot{x}^{\beta}=0$. We can then ensure that the geodesic equations obtained here are for the light trajectories, meaning that we can trace the light trajectories inside the medium for arbitrary values of $n$ including the case of $n < 1$. Following, e.g., \cite{Zhu1997} and \cite{Babar2020}, the refractive index $n$ implemented into the metric results in a non-homogenous expanding/contracting spacetime that does not violate the theory of relativity. One intuitive way to interpret this is that the null geodesics obtained for different $n$ are altered by how the spacetime is expanded or contracted due to the factor $n$, hence producing the effects of light deflection. In other words, in this context the effect of $n$ on the light deflection is equivalent to the effects of spacetime curvature on the light trajectories. Therefore, when compared between the metrics with $n<1$, $n=1$ and $n>1$, the photons emitted from the same disc location into the same direction can travel using different geodesics to the observer, producing the effects that are equivalent to the light deflection.

\subsection{Backward ray tracing}
Once the equations of motion for massless particle are derived, we can apply the standard ray-tracing technique and consider light as a photon whose trajectories follow the obtained null geodesics in the modified metric. In fact, it should be acceptable at this stage if one chooses to consider light as a photon, or vise versa, as long as we can track its geodesic. Instead of tracing emission photons from an accretion disc to a distant observer, we apply a backward ray-tracing technique to integrate photon trajectories backwards in time from the observer to the disc \citep[e.g.][]{Fanton1997, Chainakun2012, Wilkins2013}. A photographic plate (i.e. an observer's sky) is produced at a distance of $1000r_{\rm g}$ away from the black hole. The plate relating the Cartesian coordinate ($\alpha$, $\beta$) spans 4096 $\times$ 4096 pixels collecting parallel light rays from the accretion disc. Since we fix the $r_{\rm out}=50r_{\rm g}$, each pixel then corresponds to a distance of $\sim 0.02r_{\rm g}$. To map every ($\alpha$, $\beta$) to a specific radial and azimuthal position ($r_{\rm d}$, $\varphi_{\rm d}$) on the disc, the parallel rays from all pixels are traced back along geodesics using equations of motion derived in eqs.~\ref{mot1}--\ref{mot2} and eqs.~\ref{mot3}--\ref{mot4}, in an initial direction perpendicular to the photographic plate. In this way, those photons that do not reach the observer are all neglected, which save the computational time during ray-tracing simulations. 

The refractive index $n$ is switched from $n=1$ to $n<1$ or $n>1$ when the the integration reaches the radial distance of $r \leq r_{\rm cor}$. Inside the corona, the Carter constant $Q_{C}$ is changed to the value shown in eq.~\ref{Qr}. The integration of photon paths along the modified geodesics is performed until the photons, or light, intercept the accretion disc. Those that travel beyond the outer radius of the disc when $\theta > \pi/2$ and those that go inside the event horizon are neglected. The striking positions on the disc surface as well as the time taken from the observer's sky to the disc are recorded.

\subsection{Calculating emission line profiles}

The reflected flux from the accretion disc depends on the incident flux from the X-ray source. We define the reflected power per unit area as a function of radius as the disc emissivity, $\epsilon(r)$. The light bending due to the gravity of the black hole could focus light rays towards the inner disc, producing higher reflected flux from this region compared to when the space-time is flat. The emissivity profile is usually assumed to be in a power-law form varying with radius with the emissivity index $\gamma$: $\epsilon(r_{\rm d}) \propto r_{\rm d}^{-\gamma}$. Although the steeping of emissivity is varied with the geometry of the corona, the flattening of the profile to a constant emissivity index of $\sim 3$ over the outer regions is expected, regardless of the assumed geometry \citep{Wilkins2012}. 

The standard shape of the emission line can be distorted by the Doppler and relativistic effects. Photons from the accretion disc having frequency $\nu_{\rm em}$ will be observed at frequency $\nu_{\rm obs}$ by the distant observer. The redshift factor that leads to the frequency shift can be calculated via $g=\nu_{\rm obs} / \nu_{\rm em} = \bf{(p_{obs}  \cdot u_{obs}) / (p_{em} \cdot u_{em})}$, where $\bf p_{obs(em)}$ and $\bf u_{obs(em)}$ is the 4-momentum of the observed (emitted) photon and the 4-velocity of the observer (emitter), respectively. By mapping the position ($\alpha$, $\beta$) on the observer's sky to the position on the disc, the observed flux can be written as \citep[e.g.][]{Fanton1997, Chainakun2012, Wilkins2013, Jovanovic2016}
\begin{equation}
F_{\rm obs}(E_{\rm obs}) = \int\limits_{image} \epsilon(r)g^{4} \delta(E_{\rm obs} - gE_{\rm em})d\Xi ~,
\label{F_obs}
\end{equation}
where $E_{\rm em}$ is the rest energy of the emitted photon and $d\Xi$ is the solid angle subtended by the disc in the observer’s sky. The simulated line profiles are obtained by summing up the total observed flux as a function of the observed photon energies, or redshift factor, overall pixels on the photographic plate. 

%Note that the refractive index of the corona extent, $n$, can be alternatively defined through the assumed light speed inside the corona, $v_{\rm in}$.

\section{Results}
By tracing light rays along the modified Kerr geodesics, we can produce the false-colour images of accreting black holes taking into account the effects of the lensing corona. The model parameters based on our geometry setup include the source inclination ($i$), the refractive index of the corona ($n$), the coronal radius ($r_{\rm cor}$) and the emissivity index of the accretion disc ($\gamma$). Here, the false-colour images of the black hole as well as the associating line profiles varying with the model parameters are presented.

\subsection{False-colour images}
\begin{figure*}
    \centerline{
        \includegraphics[width=1.0\textwidth]{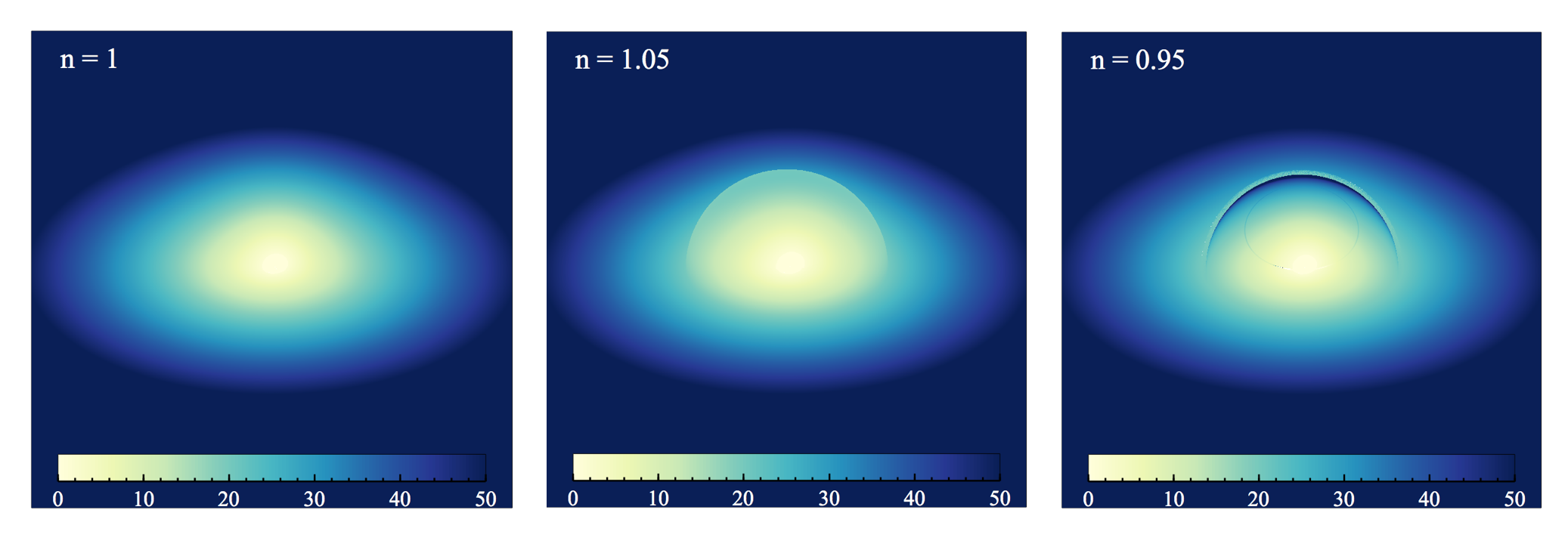}
    }
    \caption{Images of radial distance, as appeared on the photographic plate, in the unit of $r_{\rm g}$ of an accretion disc around a black hole compared between the cases when the spherical corona has the refractive index $n=1$ (left panel), $1.05$ (middle panel) and $0.95$ (right panel). We assume $a=0.998$, $i = 60^{\circ}$ and $r_{\rm cor}=20r_{\rm g}.$ Lensing effects due to the presence of the corona with $n \neq 1$ show an influence on the appearance of the accretion disc compared to the standard case of the corona when $n=1$.      
    \label{img-r-n}}
\end{figure*}

First of all, we investigate how the refractive index of the corona affects the radial map of the accretion disc as appeared on the photographic plate. Fig.~\ref{img-r-n} shows the false-colour images of the radial distance measured in the unit of $r_{\rm g}$ of the accretion disc around an extreme Kerr black hole ($a=0.998$) viewed at an inclination angle of $i = 60^{\circ}$. We freeze the parameters $r_{\rm cor}=20r_{\rm g}$ and $\gamma=3.0$. The refractive index of the corona is varied to be $n=0.95$, 1, 1.05. The refractive index $n$ included in the modified metric alters the null geodesics followed by light, hence produces the effect which is analogous to the light deflection. It can be seen that the presence of a lensing corona affects the appearance of the accretion disc, especially on the far side within the corona.

\begin{figure}
    \centerline{
        \includegraphics[width=0.45\textwidth]{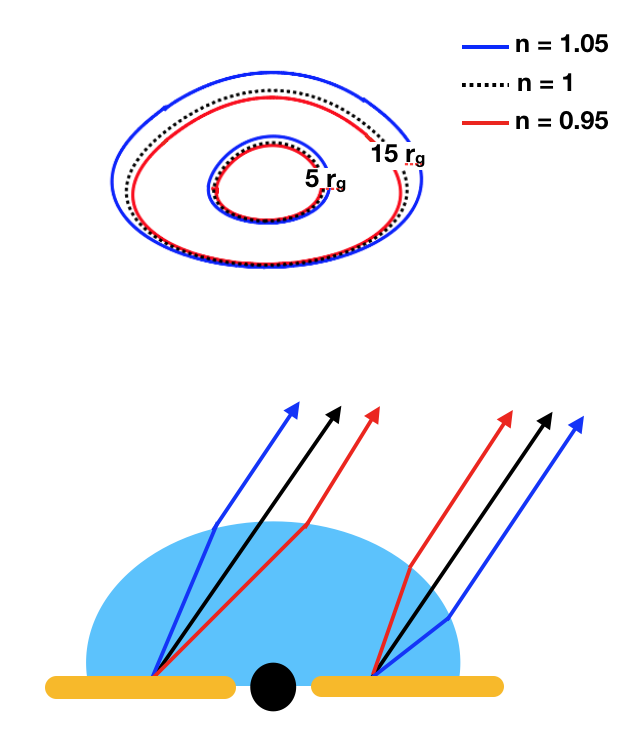}
    }
    \caption{Top panel: Contours at radii of 5 and 15$r_{\rm g}$ for different $n$ as can be seen on the images of the radial distance of the disc presented in Fig.~\ref{img-r-n}. Due to Doppler and relativistic light-bending effects, the contour lines appear to be compressed in the near side of the hole. Moreover, the corona with $n>1$ and $n<1$ produces the image of the inner disc that appears further away and closer in, respectively. Bottom panel: An illustration to explain this phenomenon (not to scale). The parallel light rays that reach different pixels of the photographic plate for different values of $n$ can be traced back to the same location on the disc. This produces the contour lines that are relatively further in when $n<1$ and relatively further out when $n>1$ due to different light deflections at the coronal boundary. See text for more details. 
    \label{contour_r}}
\end{figure}

Corresponding contour lines on the disc at radii of 5 and 15$r_{\rm g}$ are shown in Fig.~\ref{contour_r} (top panel). When the corona has $n>1$, it can be understood that the light ray is bent away from the normal line upon crossing the coronal boundary from inside (denser medium) to the empty space outside (vacuum, $n=1$). This is why the disc on the far side within the corona, as seen by the observer, appears on the photographic plate at a larger distance compared to the case of the corona with $n=1$. On the other hand, when $n<1$, the light is bent toward the normal line upon crossing the coronal shell, producing the disc images whose contour lines are relatively further in. This phenomenon is illustrated in Fig.~\ref{contour_r} (bottom panel). The different appearances of the accretion disc then affect the observed photon flux and the calculated emission lines. 
  
\begin{figure*}
    \centerline{
        \includegraphics[width=1.0\textwidth]{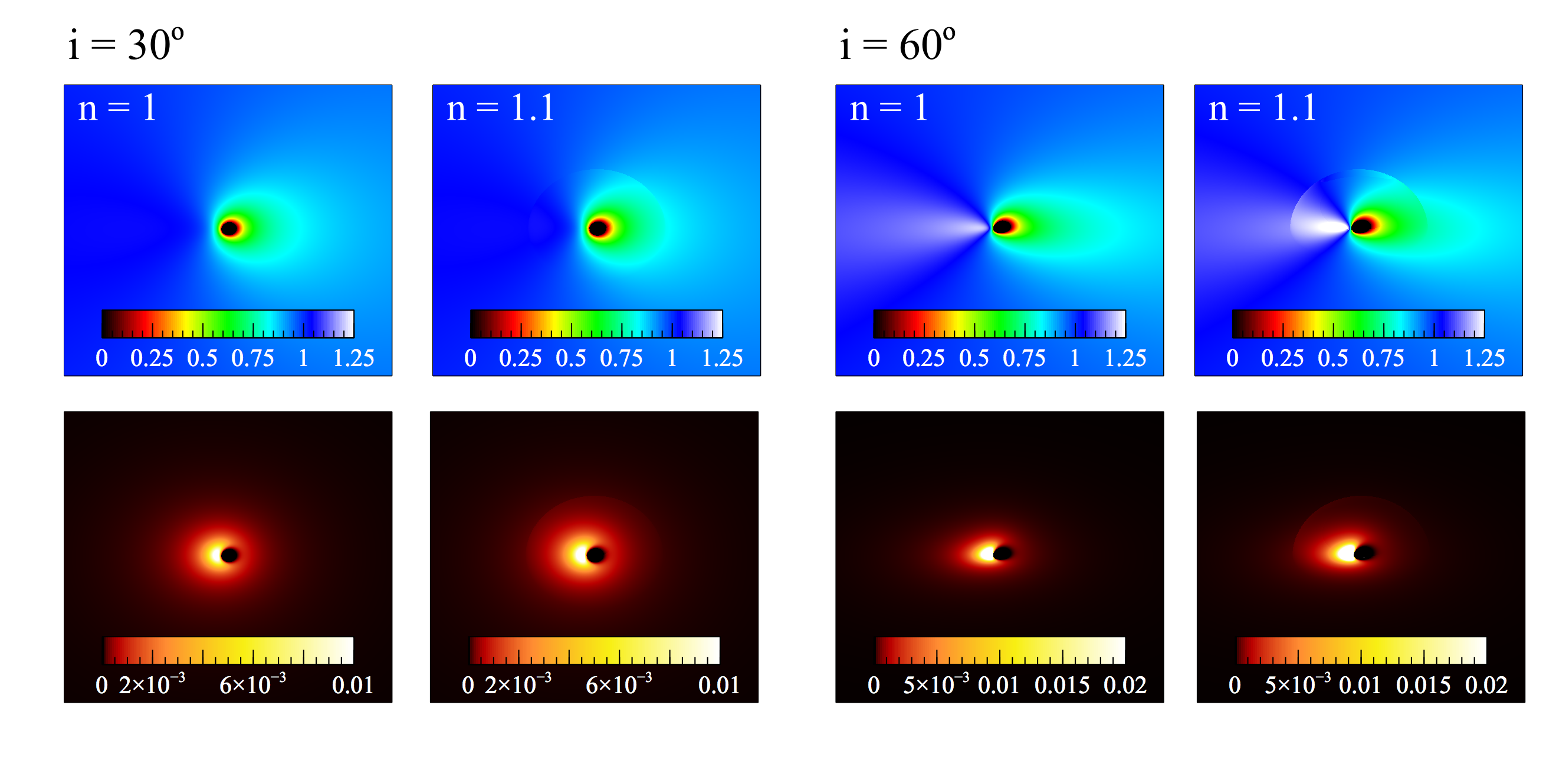}
    }
    \vspace{-0.5cm}
    \caption{Images of the frequency shift (upper panels) and the corresponding observed flux (bottom panels) of the accretion disc for the inclination angles of $30^{\circ}$ and $60^{\circ}$ compared between the cases when the effect of lensing corona is absent ($n = 1$) and present ($n = 1.1$). Other parameters are $a=0.998$, $\gamma=3.0$ and $r_{\rm cor} = 20r_{g}$. The most intense emission is from the left-hand side of the image corresponding to the regimes where the accretion disc is moving towards the observer.      
    \label{fig-img-g-f}}
\end{figure*}

Fig.~\ref{fig-img-g-f} (top panels) shows the false colour images of the redshift factor, $g$, of the photons from the different parts of the accretion disc as seen by the observer at $i=30^{\circ}$ and $i=60^{\circ}$. We compare, as an example, the cases of $n=1$ and $n=1.1$. The size of the corona is fixed at $r_{\rm cor} = 20r_{\rm g}$ and the disc has the emissivity index of $\gamma =3$. The left and right parts of each image indicate the frequency shift of emitted photons from the approaching and receding sides of the disc, respectively. The bluer image (i.e. larger blueshift) could be seen in higher inclination cases. This is expected since the observer with higher inclination should experience more Doppler boosting caused by the motion of the accreting gas with respect to the line of sight. Furthermore, the presence of the lensing corona with $n>1$ could enhance the blueshift of those photons emitted from the accreting gas within the corona, especially from the gas that is moving towards us. This effect is more significant for higher inclinations. The corresponding observed flux images are presented in Fig.~\ref{fig-img-g-f} (bottom panels). Due to the Doppler beaming effects, the apparent brightness of the accretion disc that is moving towards (away from) the observer will appear relatively brighter (fainter). The lensing corona can further amplify the observed flux because the photon flux in the observer's frame is modified by the larger factor of $g^{4}$, as expressed in eq.~\ref{F_obs}.  

Once the redshift and the observed flux images are obtained, the emission line profiles can be produced by summing up the flux overall disc elements with the same range of the redshift factor. From now on, the emission lines simulated in the case of the corona with $n=1$ are always presented in the dotted lines. Note that when $n=1$, the simulated emission lines should have the standard asymmetric profiles with bright blue peak and extended red tail which were extensively studied before in previous literature \citep[e.g.][]{Fabian2000, Reynolds2003}. If not specified, the model parameters are fixed at $r_{\rm cor}=20r_{\rm g}$, $n=1.05$, $i=30^{\circ}$ and $\gamma=3$.

\subsection{Inclination}
\begin{figure}
    \centerline{
        \includegraphics[width=0.45\textwidth]{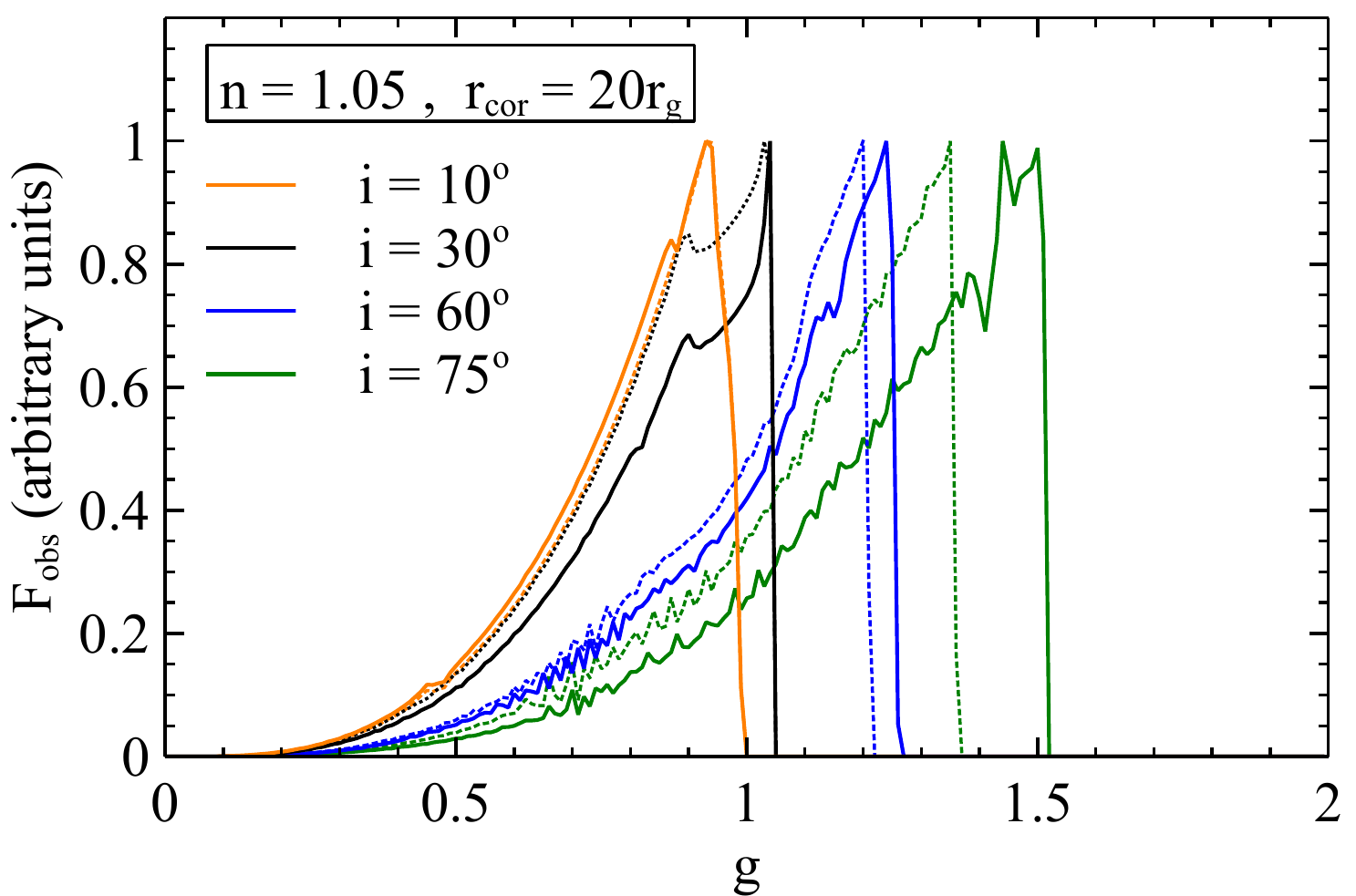}
    }
    \vspace{0.2cm}
    \centerline{
        \includegraphics[width=0.45\textwidth]{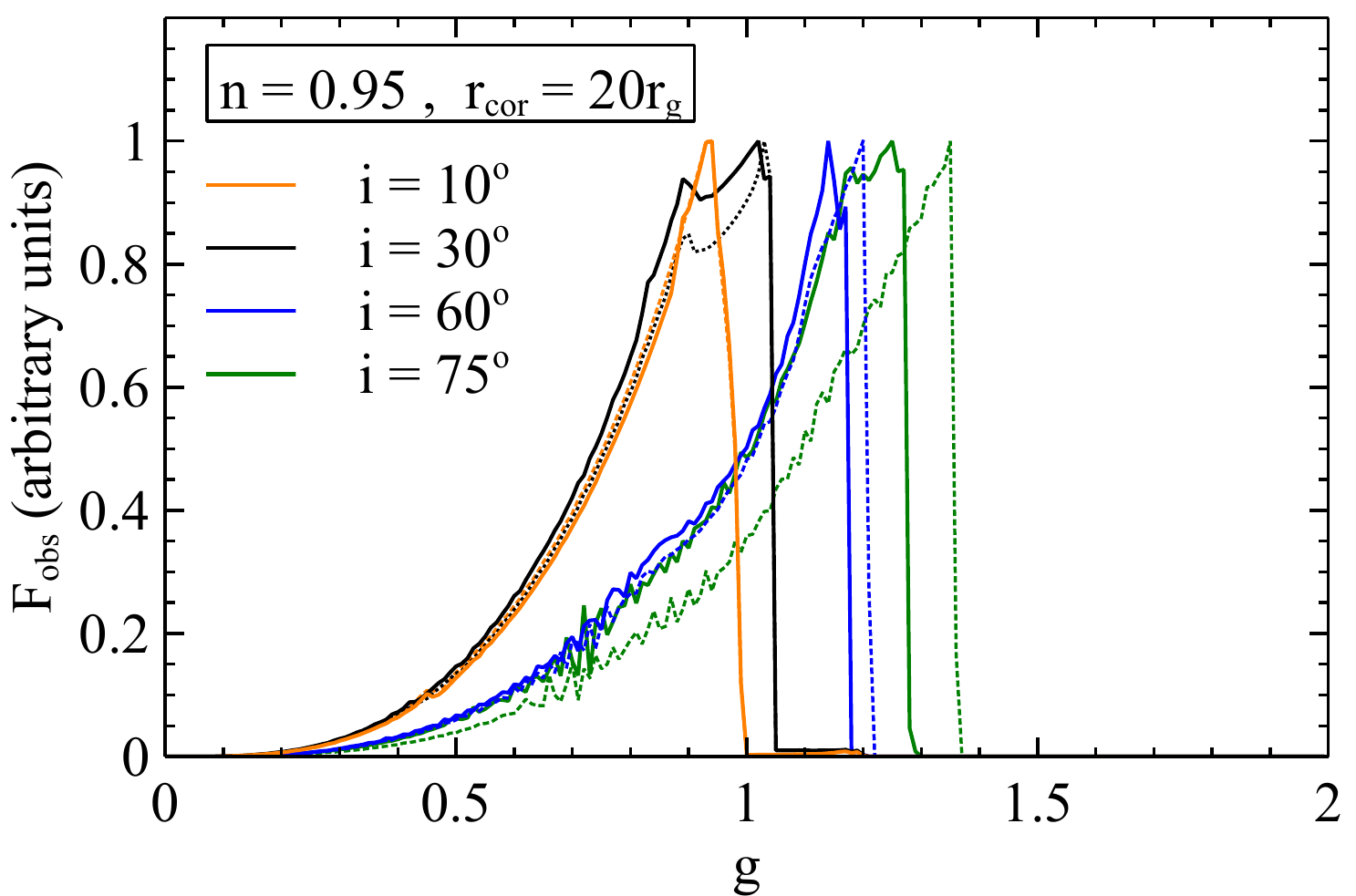}
    }
    \caption{Emission line profiles of an accretion disc around an extreme Kerr black hole for the inclination angles of $i=10^{\circ}$, $30^{\circ}$, $60^{\circ}$ and $75^{\circ}$ showing in the orange, black, blue and green solid lines, respectively, when the corona has the refractive index of $n=1.05$ (top panel) and $n=0.95$ (bottom panel). We assume the coronal size of $20r_{\rm g}$ and the emissivity index of the disc is $\gamma = 3$. For comparison, we show in the dotted lines the corresponding cases when we treat the corona as an optical medium with $n=1$.}
    \label{fig-line-i}
\end{figure}

Effects of the inclination on the emission line profiles are shown in Fig.~\ref{fig-line-i}. For comparison, we normalize the highest flux in each emission line to 1. The observer at high inclinations faces much stronger effects of Doppler shift and relativistic beaming than the one at low inclinations (i.e. the blue peak appears at higher $g$ for higher inclination). While the overall shape and the trend of the lines are in agreement with what reported in previous literature \citep[e.g.][]{Fanton1997}, we can observe that the blue wing of the line is further shifted to higher $g$ when the corona is treated as a lensing object with $n>1$ (Fig.~\ref{fig-line-i}, top panel). This effect is more dominant for higher inclinations. In fact, for the lensing corona with $i=75^{\circ}$, we observe the highest blueshifted photons at $g \sim 1.5$, while those found in the standard corona case are maximally at $g \sim 1.4$. The shapes of the line profiles obtained here are analogically similar to what can be implied from the redshift and flux images presented in Fig.~\ref{fig-img-g-f}.

Contrarily, Fig.~\ref{fig-line-i} (bottom panel) shows that if the corona has $n<1$, the extra redshift, rather than blueshift, is seen. This is in agreement with the different appearances of the disc image presented in Fig.~\ref{contour_r}. The effect of the line shift due to the presence of the lensing corona is also more drastic for higher inclinations.

\subsection{Refractive index}
\begin{figure}
    \centerline{
        \includegraphics[width=0.45\textwidth]{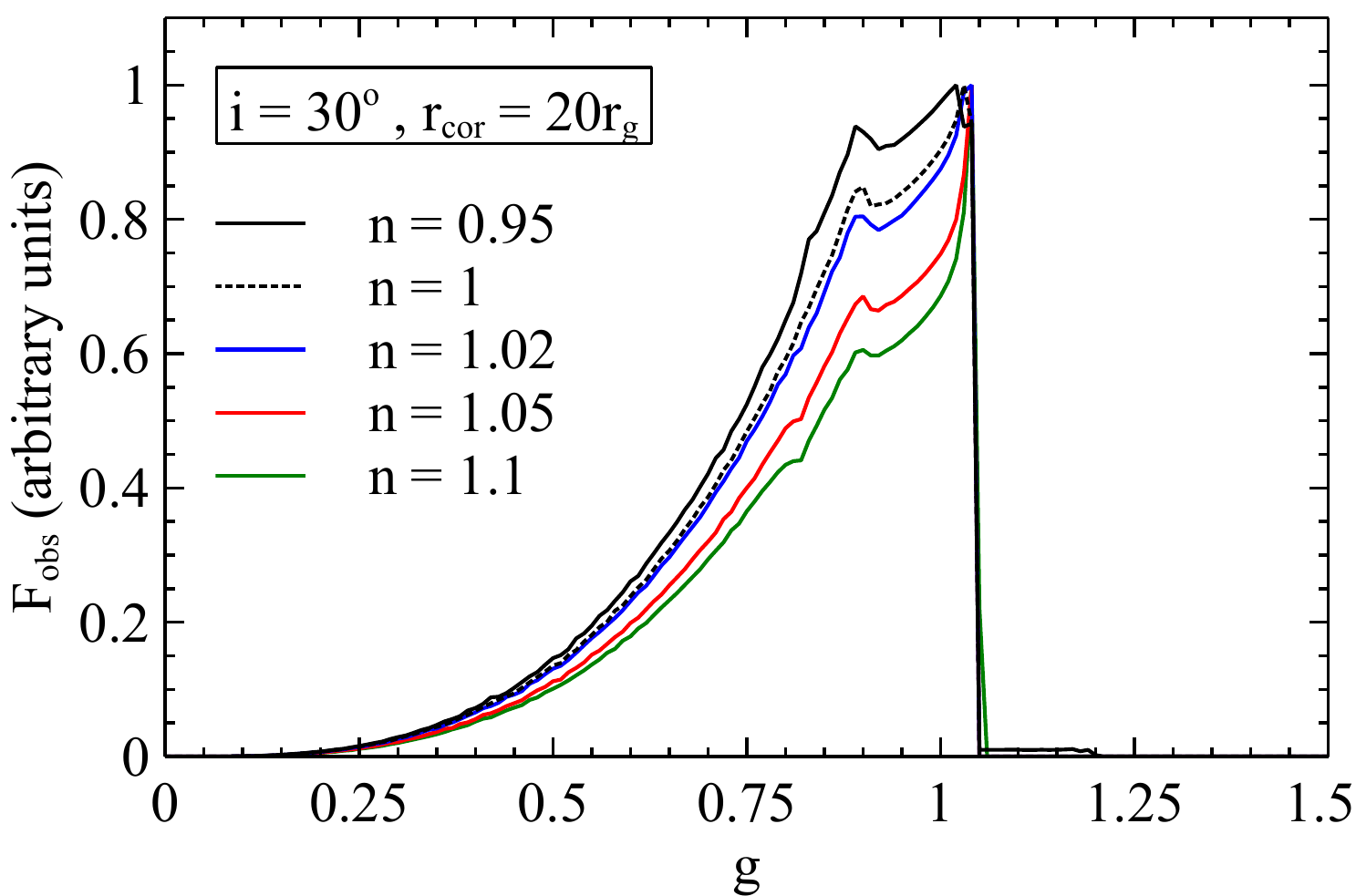}
    }
    \vspace{0.2cm}
    \centerline{
        \includegraphics[width=0.45\textwidth]{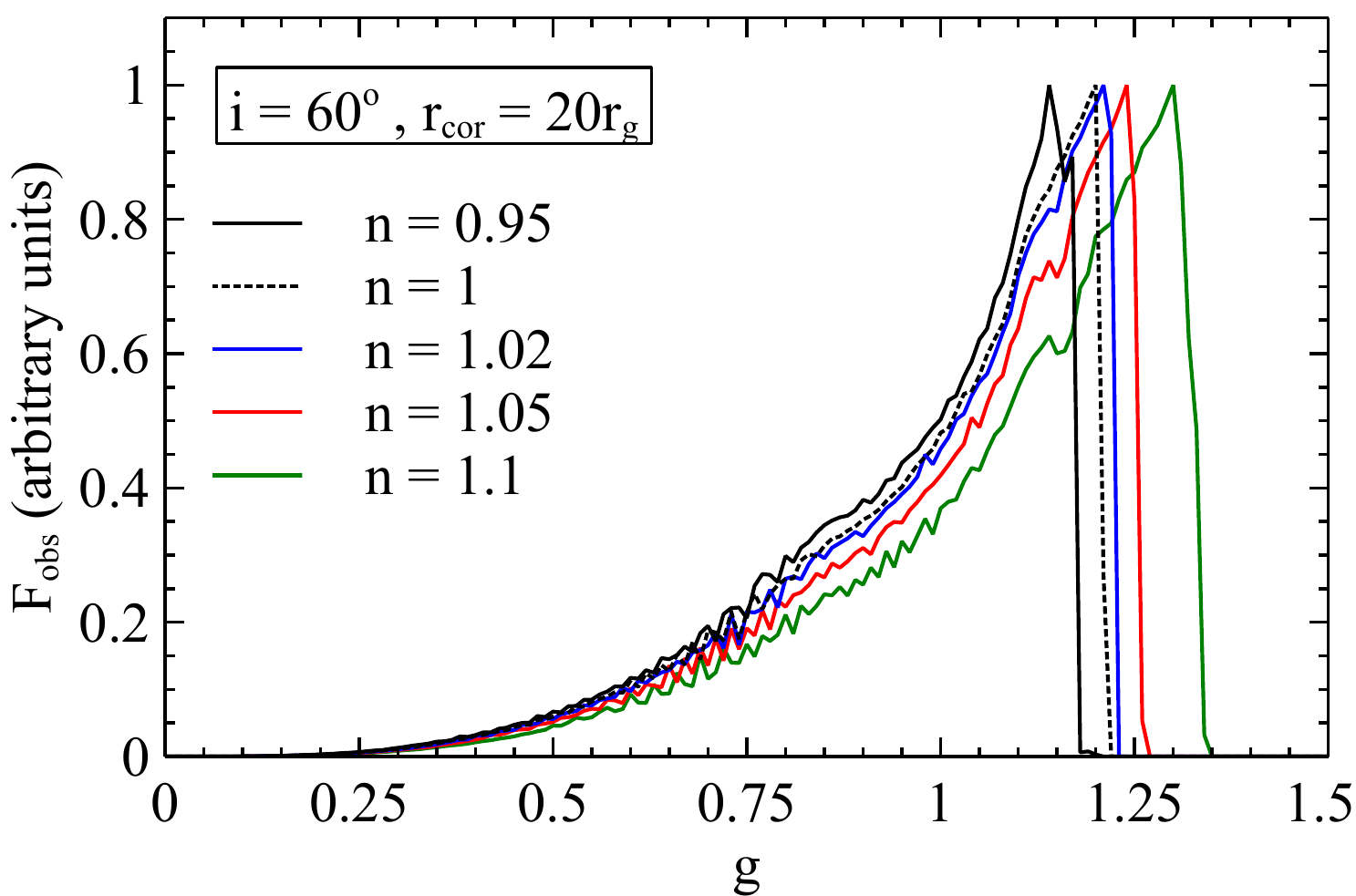}
    }
    \caption{Emission line profiles varying with the refractive index $n$ when $i=30^{\circ}$ (top panel) and $60^{\circ}$ (bottom panel). We assume $r_{\rm cor} = 20r_{\rm g}$ and $\gamma = 3$.}
    \label{fig-line-i2}
\end{figure}

We now consider the general effects of $n$ on the line profiles. Fig.~\ref{fig-line-i2} shows that the corona with more difference of $n$ from 1 leads to the observed line profiles that are more deviated from the standard emission lines. Higher $n$ means that the corona is denser and the light wave coming from the far side inside the corona is more bent, at the coronal boundary, further away from the normal line. The photons from the inner regions, most of which are responsible for the blue wing, then cover a larger radial distance as appear on the photographic plate, inducing more flux difference between the blue peak and those contributed to the extended red tail. As a result, the blue peak is more enhanced with respect to the red tail. For higher inclination ($i=60^{\circ}$), larger Doppler boosting effects widens the line shapes by shifting the blue peak to higher $g$. When the corona is denser, the emission photons from the inner regions are shifted to higher energy, hence the line shapes are further broadened. On the other hand, when $n<1$, more redshifted photons are observed and the line is shifted towards the red sides narrowing the line profile. The effects of broadening and narrowing the lines due to the lensing corona with $n>1$ and $n<1$, respectively, are more significant for higher inclinations.

\subsection{Coronal size}
\begin{figure}
    \centerline{
        \includegraphics[width=0.45\textwidth]{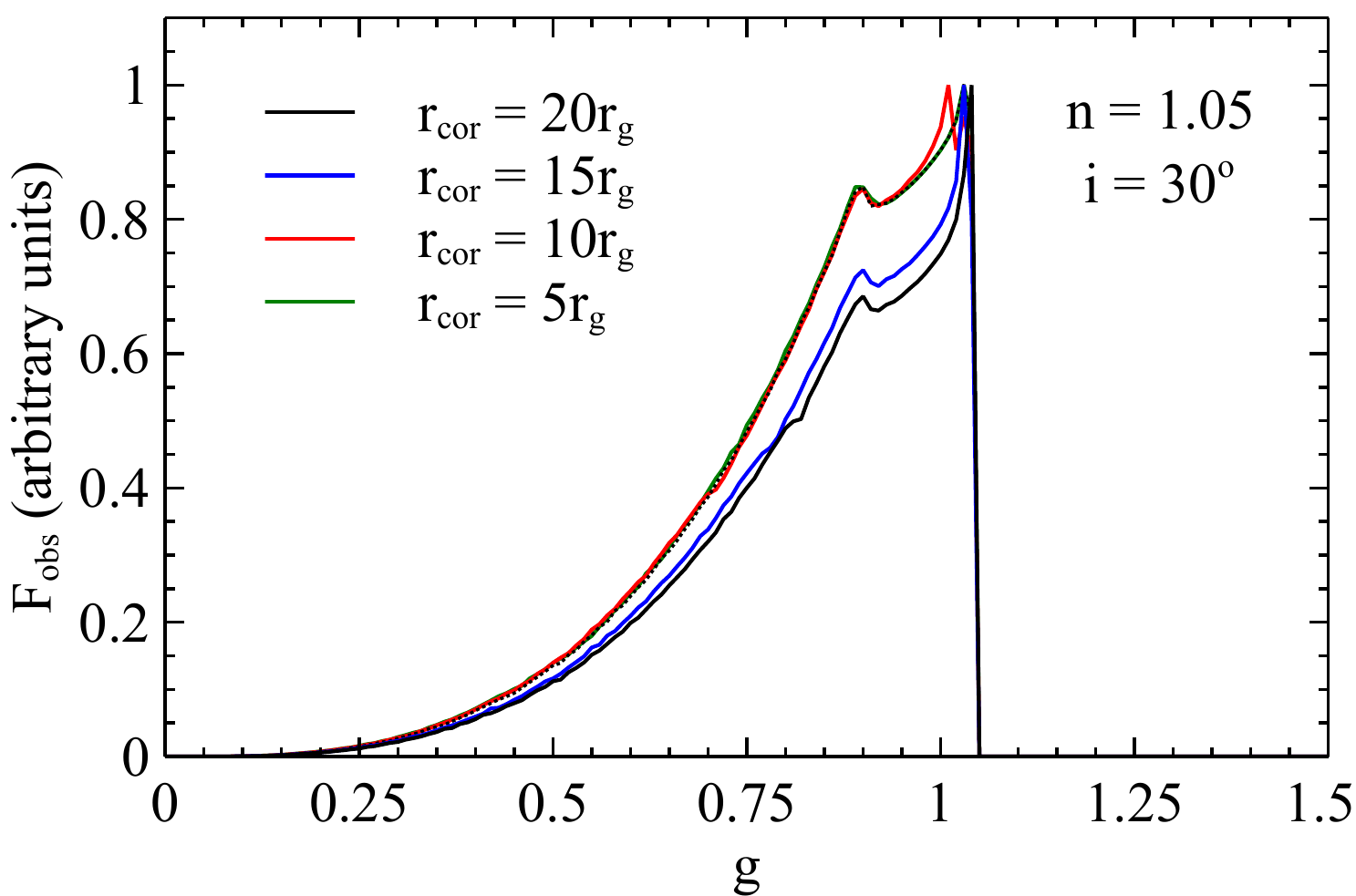}
    }
    \caption{Emission line profiles varying with $r_{\rm cor}$ when $n=1.05$, $i=30^{\circ}$ and $\gamma = 3$. For comparison, we present in the dotted line the case when $n=1$. The small size of the corona such as $5r_{\rm g}$ could almost resemble the case that the effects of coronal lensing are ignored.
    \label{fig-line-rcor}}
\end{figure}

Next, we consider the effects of the coronal size. Fig.~\ref{fig-line-rcor} shows the dependence of line shapes on the coronal radii of 5, 10, 15 and 20$r_{\rm g}$. All the models have $n=1.05$, $a=0.998$, $i=30^{\circ}$ and $\gamma=3$. It can be seen that decreasing the coronal size decreases the effect of coronal lensing, which is expected since the number of light rays that cross upon the coronal boundary decreases. For $r_{\rm cor}$ that is as small as $5r_{\rm g}$, the shape of the line profile almost resembles that produced when $n=1$ (i.e. the green solid line and the black dotted line in Fig.~\ref{fig-line-rcor} are almost identical). Although the results of the corona with $n<1$ are not shown here, less deviation of the emission line is expected for smaller $r_{\rm cor}$.

\subsection{Disc emissivity}
\begin{figure}
    \centerline{
        \includegraphics[width=0.45\textwidth]{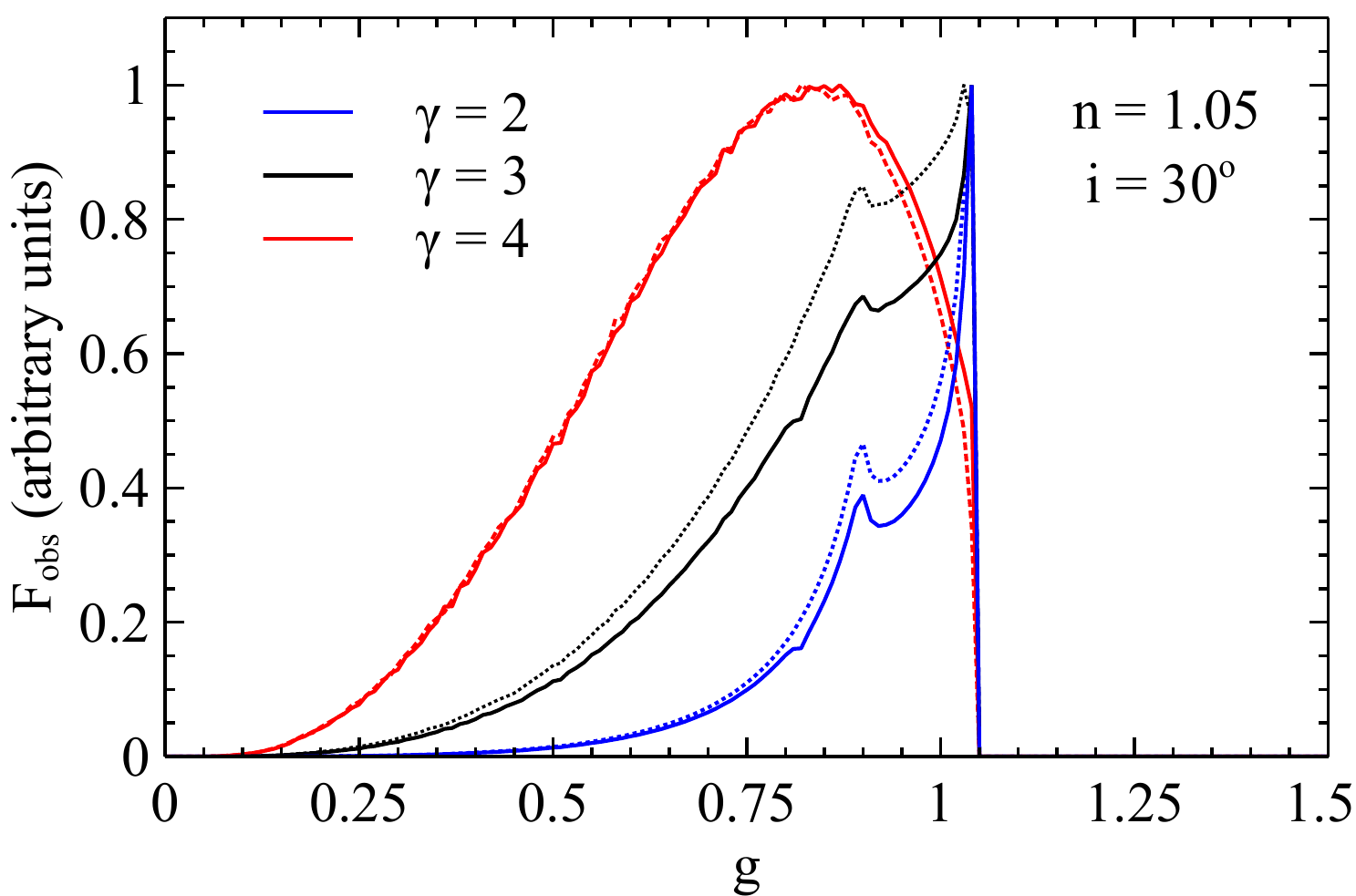}
    }
    \caption{Emission line profiles varying with emissivity index $\gamma$ when we fix $n=1.05$, $r_{\rm cor}=20r_{\rm g}$ and $i=30^{\circ}$. Dotted lines represent the corresponding profiles when $n=1$.
    \label{fig-line-index}}
\end{figure}

Effects of the disc emissivity on the emission line profiles are presented in Fig.~\ref{fig-line-index}. Higher $\gamma$ means the radiation is more concentrated around the centre. The emissivity index is then the parameter that regulates the importance of emission across all disc radii. Note that $\gamma$ does not depend solely on the coronal geometry, but also on, e.g., the ionisation state of the disc \citep[see][and discussion therein]{Kammoun2019}. Therefore, we do not tie the emissivity index to the geometry here, but instead allow it to be a free parameter in order to investigate how it could affect the deviation of the line profiles in general. Varying $\gamma$ results in the changes in relative flux difference between the blue wing and the extended red tail compared between the case when $n=1$ and $n=1.05$. The deviation of the lines is still noticeable in the range of $\gamma \sim$~2—4 even for a low inclination of $i=30^{\circ}$, as evidenced in the solid and dotted lines in Fig.~\ref{fig-line-index}.

\section{Discussion}

The modified Kerr metric and corresponding equations of motion are derived to trace the light trajectories along geodesic lines taking into account the effects of the light-bending caused by gravity and the lensing corona. The combination of these effects produces the line profiles with greater precision. The coronal shell acts as an interface that separates two optical media with different refractive indices: vacuum ($n=1$) and the coronal extent ($n \neq 1$). The modified metric produces light deflection by altering the null geodesics dependent on $n$, hence affects the trajectories of light rays that pass through different medium (vacuum and corona). According to our results, these lensing effects are more important when the coronal size is large or when the refractive index is more different from 1. Larger $i$ also induces more deviation of the emission lines from those produced in the standard corona case of $n=1$.

We assume the disc emissivity in the form of a simple power-law with an emissivity index $\gamma$. \cite{Wilkins2012} showed that if the coronal height is low, emissivity is centrally concentrated (large $\gamma$) and a very broad emission line could be produced. Higher source height leads to smaller $\gamma$ and narrower emission lines because of relatively small differences in the reflected flux across all radii. Here, we show that in an extended corona case, the broadening and narrowing of the lines could be induced by the changes in the refractive index of the corona itself, apart from the changes in the coronal geometry. 

Due to the great effects of the radial coronal extent, emissivity profiles produced by the point source and spherical corona are distinguishable without requiring an analysis of reflection fraction \citep{Gonzalez2017}. The corona extended radially produces the intense incident flux in the region below the source, inducing the disc emissivity to be flatter over the middle part. The emissivity profile specific to the spherical coronal geometry then can be in a more complex form than what is employed in this investigation. However, \cite{Kammoun2019} showed that steeper emissivity profiles could be obtained by allowing the radial gradient of the disc ionisation, and the effects due to the radial ionization profile may be more prominent than the geometrical effects. Subject to this uncertainty, we do not convey further the complex shape of the emissivity profiles, but instead point out that the interplay between the coronal geometry, disc ionisation state, and refractive index of the corona could possibly contribute to the formation of the line shape. The one that plays the most important role may differ among different AGNs. 

The most notable effect of the lensing corona is the additional shift of the lines towards higher energy and lower energy when $n>1$ and $n<1$, respectively (Fig.~\ref{fig-line-i}). Different geometry set-up such as an accretion disc with a finite thickness \citep{Taylor2018} or a disc with a high-density \citep{Jiang2018} could produce their own emission line profiles. We then expect degeneracies of these emission lines produced by different frameworks. For example, \cite{Wilkins2020} studied the reflected X-rays that return to the accretion disc due to the strong gravity of the black hole. They found that the effects of returning radiation could enhance the observed iron K$\alpha$ line and slightly redward the peak of the line. This extra redshift of the line is analogically similar to what is produced by the lensing corona with $n<1$. Breaking these degeneracies may not be straigthforward, so we choose to focus only on this lensing-corona scenario and investigate further on the quantitative differences of our obtained line shifts. 

\begin{figure}
    \centerline{
        \includegraphics[width=0.45\textwidth]{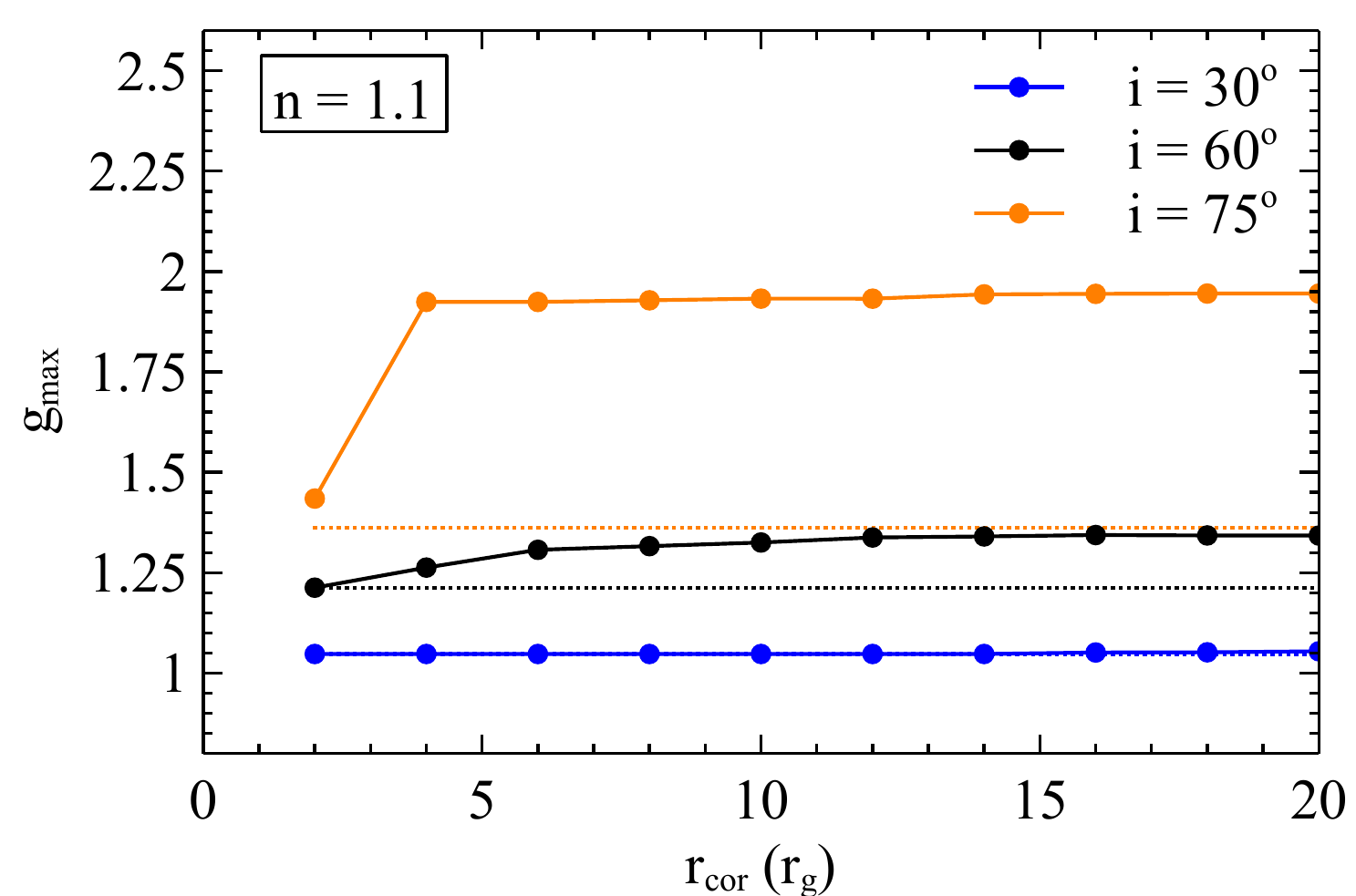}
    }
    \vspace{0.2cm}
    \centerline{
        \includegraphics[width=0.45\textwidth]{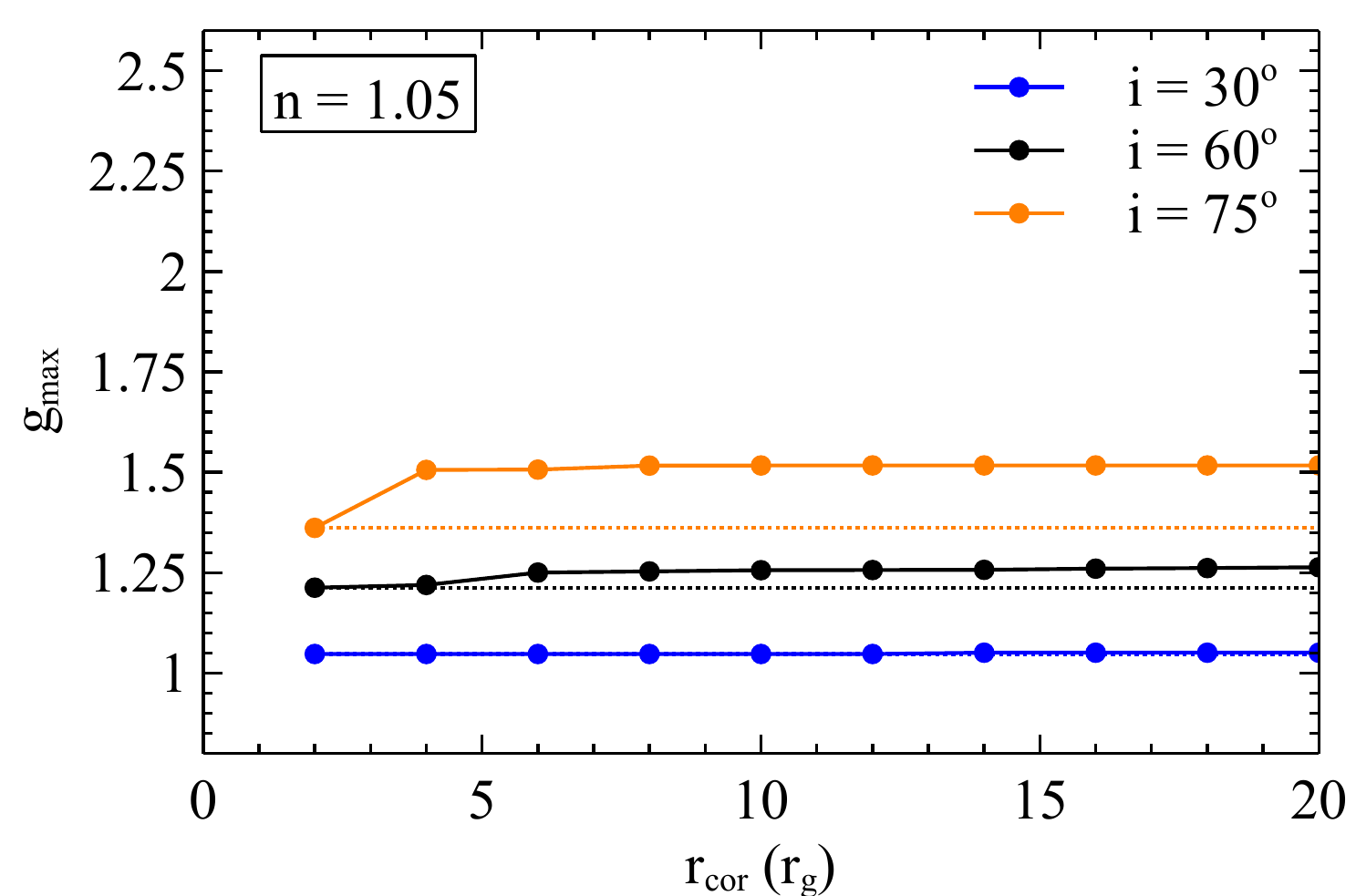}
    }
    \caption{Maximum frequency shift where the blue peak of the emission line is observed varying with the coronal size and inclination when $n=1.1$ (top panel) and $n=1.05$ (bottom panel). Dotted lines represent the corresponding cases when $n=1$. }
    \label{fig-gmax}
\end{figure}

\begin{figure}
    \centerline{
        \includegraphics[width=0.45\textwidth]{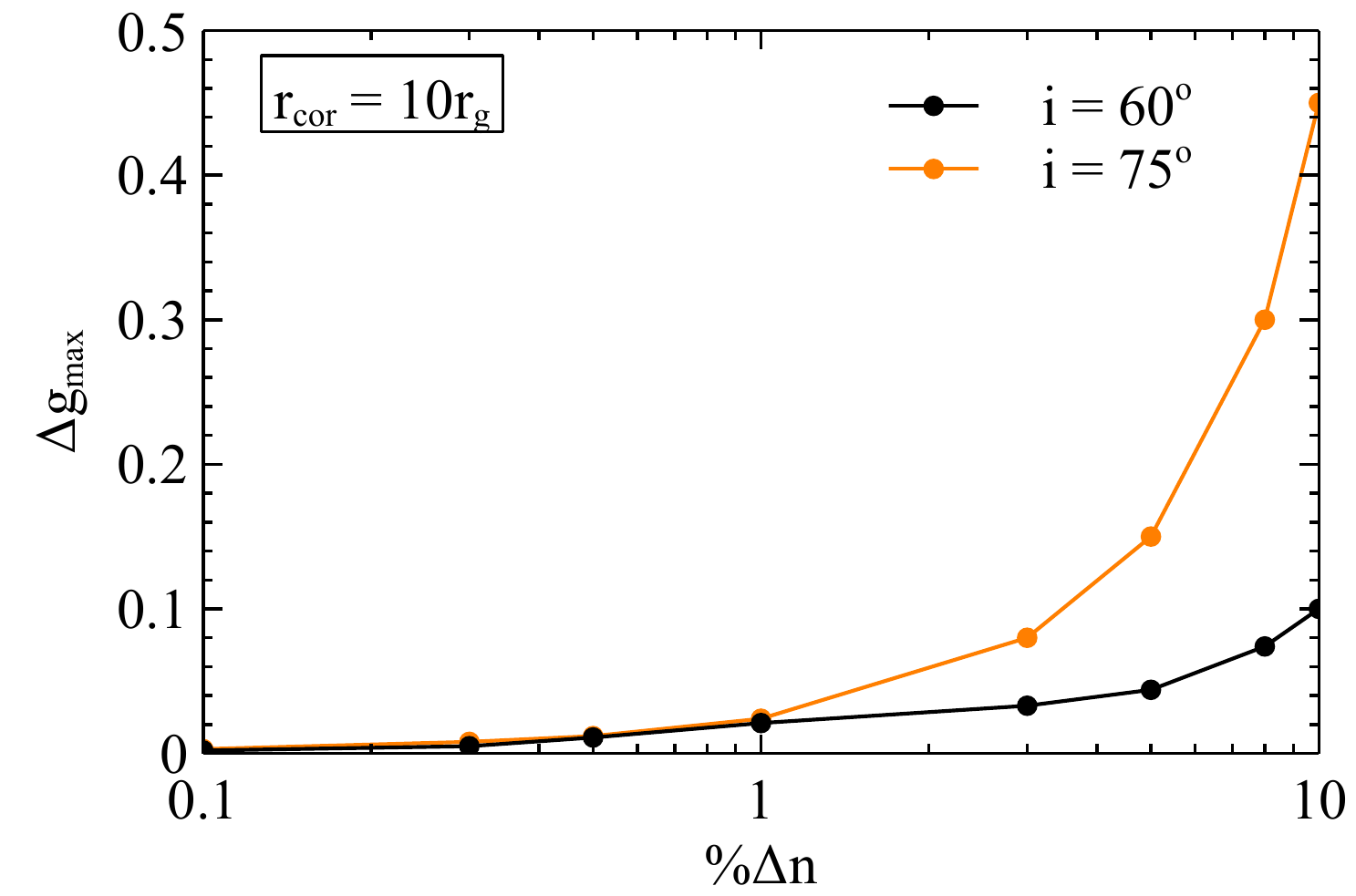}
    }
    \caption{Extra line shift ($\Delta g_{\rm max}$) varying with the percentage difference between the refractive index of the corona and that of the vacuum ($\% \Delta n$) when $r_{\rm cor}=10r_{\rm g}$. The black and orange lines correspond to the case when $i=60^{\circ}$ and $75^{\circ}$, respectively. To obtain $\Delta g_{\rm max} \gtrsim 0.01$, it requires $\% \Delta n \gtrsim 0.5$.
    \label{fig-delta-gmax}}
\end{figure}

Fig.~\ref{fig-gmax} shows how the frequency that the blue peak appears depends on the coronal size. The strongly blueshifted lines up to $g_{\rm max} \sim1.9$ is found when $i=75^{\circ}$ and $n=1.1$. This effect seems to be less dependent on the coronal size as long as $r_{\rm cor} \gtrsim 5r_{\rm g}$ which is large enough to cover the inner accretion disc and enhance the energy shifts of the strongly blueshifted photons originated from these regimes. The additional blueshift is subtle either when $n$ is closer to 1 or when the inclination angle is smaller. Perhaps, the effect of lensing corona to the emission lines could be probed by identification of the extra line shifts of the blue wing in the AGN observed with $i \gtrsim 60^{\circ}$. Fig.~\ref{fig-delta-gmax} represents the extra shift of the line ($\Delta g_{\rm max}$) varying with the percentage difference between the refractive index of the corona and the vacuum ($\% \Delta n$). To obtain $\Delta g_{\rm max} \gtrsim 0.01$, the percentage difference of the refractive index of the corona and that of the vacuum needs to be $\% \Delta n \gtrsim 0.5$ (i.e., $n\gtrsim 1.005$ or $n\lesssim 0.995$). When $\% \Delta n$ increase, $\Delta g_{\rm max}$ increases more significantly for higher $i$. To obtain $\Delta g_{\rm max}$ up to $\sim 0.1$, it requires the deviation of the coronal refractive index of $\sim 3.5\%$ and $10\%$ from that of the empty space for $i=75^{\circ}$ and $60^{\circ}$, respectively. Note that the similar trend but with extra redshift would expect in cases of $n<1$.

\begin{figure}
    \centerline{
        \includegraphics[width=0.45\textwidth]{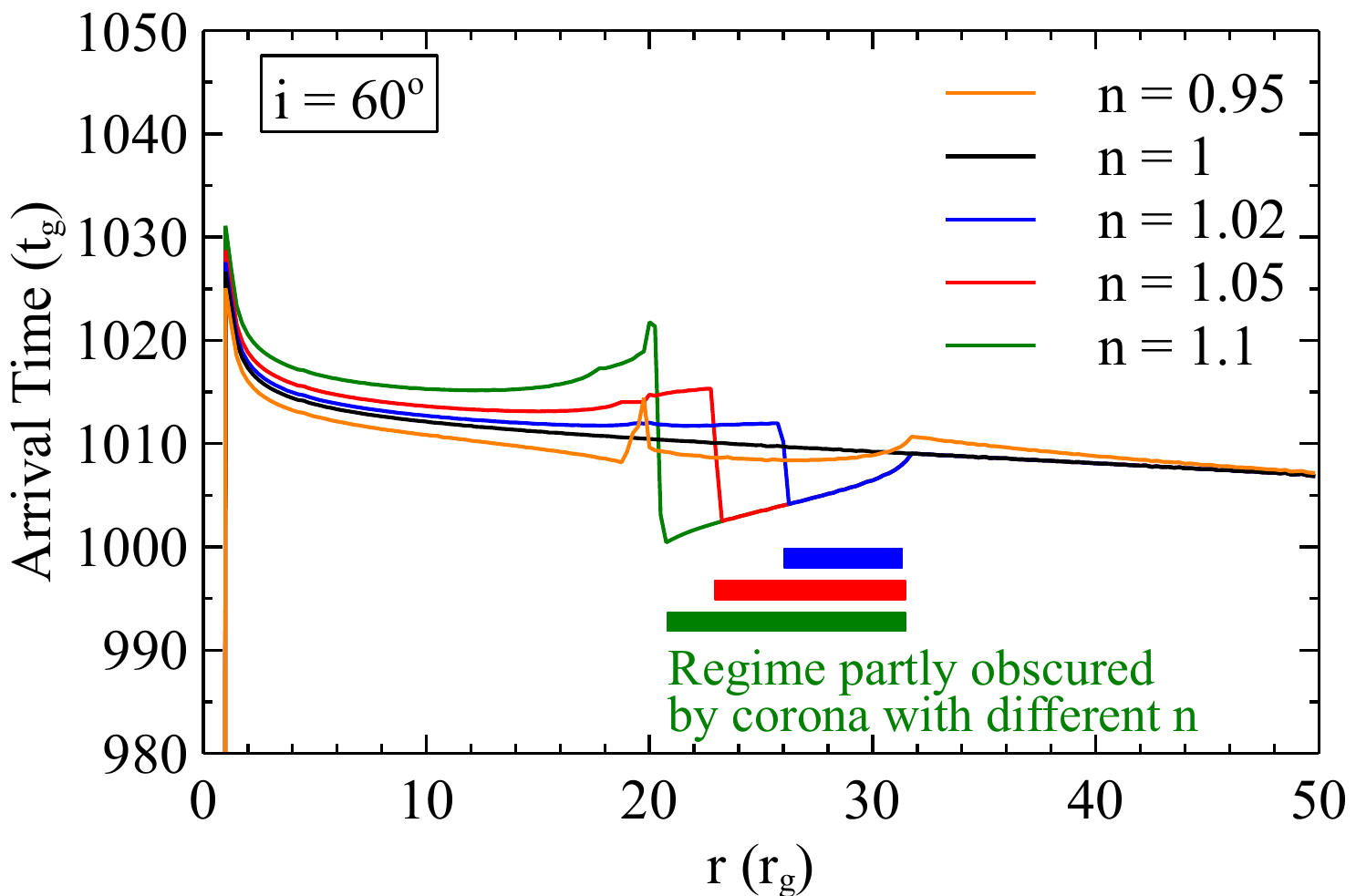}
    }
    \caption{Average arrival time of photons from different radii of the accretion disc varying with $n$ when $r_{\rm cor}=20r_{\rm g}$, $i=60^{\circ}$, and $\gamma = 3$. The coloured bands represent the area where the corresponding profile shapes are manifested due to some parts of the disc are obscured by the corona. See text for more details.}
    \label{fig-t}
\end{figure}

As another application, we show in Fig.~\ref{fig-t} the average arrival time of photons from different disc radii to the observer's sky located at $1000r_{\rm g}$ away from the central black hole. For $n>1$, the lensing corona increases the averaged light-travel time of photons originated from the inner disc, meaning that the light-travel paths between the inner disc and the observer are longer compared to the case when $n=1$. This highlights the effects of the light deflection at the coronal boundary induced by $n$ (light takes different paths for different $n$). However, the far side of the disc outside the corona ($20r_{\rm g}<r_{\rm cor}<30r_{\rm g}$ in this case) is obscured by the corona extent, hence at these regions the observer sees mostly the photons from the near side, sampled out those photons taken relatively long light-travel time from the far side. As a result, the average time taken by photons travelling from these radii to the photographic plate decreases, producing a dip in the profiles. 

The switch from longer to shorter arrival time when $n>1$ occurs at smaller radii for larger $n$. This is because light rays travelling from specific disc radii on the far side outside the corona are supposed to be bent once intercept the corona shell. Increasing $n$ in this configuration leads to more light bending away from the normal line, so the parallel light rays incident on the same pixel are traced back to smaller radii for larger $n$, and as a result, the obscured regime extends more inwards. On the other hand, in the case of $n<1$ (Fig.~\ref{fig-t}, orange line), the observed arrival time of those photons that depart from the inner disc is shorter, rather than longer. We note that one way to intuitively think about the effects of the refractive index on the arrival time is that the null geodesics change when we change the metric. The photons then take different paths to the observer in modified and non-modified Kerr spacetime. This is why it is likely that some photons travel faster than $c$ (arrive earlier compared to the case of the Kerr metric), but they actually just take different, shorter paths in the modified geodesics. According to these results, the lensing corona with $n<1$ produces inverse effects on the timing profiles as well as on the time-averaged profiles discussed before when compared to the case of $n>1$. 

While the presence of the lensing corona could affect the timing profiles as well as the mean spectra, the photons emitted from the disc radii further than $r_{\rm cor}$ are less affected under this scenario. Although we fix the outer edge of the disc at $r_{\rm out} = 50r_{\rm g}$, increasing $r_{\rm out}$ should not change the comparative differences of the simulated data in the corresponding cases of the corona with $n=1$ and $n \neq 1$, as long as the coronal size is significantly smaller than the size of the accretion disc.

There is mounting evidence that the extended corona is preferable over the static lamp-post configuration in explaining the observed X-ray timing data in AGN \citep[e.g.][]{Caballero2018, Chainakun2019b, Zoghbi2021}. The shapes of extended corona could be a slab, sphere, or even outflowing jet, and could also change over time. For simplicity, we employ the spherical corona, but it could represent a compact state of the corona that is confined in a spherical region after being collapsed \citep{Wilkins2015, Gallo2018}. We restrict the maximum radius of the corona at $20 r_{\rm g}$, which is likely to be the maximum size of the extended corona obtained using different techniques \citep[e.g.][]{Reis2013, Adegoke2017, Chainakun2019b}.

While our chosen geometry is specific, the method presented here is general in the way that it should be applied to other extended coronal geometries. In this work, the study is simplified to the case of the corona whose optical properties are specified by a constant refractive index $n$. The model can be further improved in several ways such as by allowing the coronal shell to have a finite thickness and apply a smooth monotonic increasing function to implement the change in the refractive index through the shell. The exact value of $n$ of the AGN corona is still elusive. Furthermore, for relativistic electron gas, the refractive index can be greater or smaller than $1$, or even taken as negative values \citep{deCarvalho2016}. The optical properties of corona are usually explained in previous literature using the optical depth, but the reported values are quite scattered. The optical depth for the AGN corona obtained by fitting the broad-band spectra is found to be 10--40 \citep{Petrucci2018}. An optically thick corona with the optical depth of 2--10 was also implied by fitting the lag-frequency profiles \citep{Chainakun2019b}. Nevertheless, it was argued by, e.g., \cite{Steiner2017} that the corona should be optically thin, otherwise the reflection features will be smooth out when passing through the corona before being observed. 

Ones can relate the optical depth, $\tau$, to the number density of the coronal electrons, $N_{\rm e}$, via $N_{\rm e} = \tau/(\sigma_{\rm T}r_{\rm cor})$, where $\sigma_{\rm T}$ is the Thomson scattering cross-section and $r_{\rm cor}$ is the radius (size) of the corona \citep[e.g.][]{Inoue2021}. Treating the coronal electrons with the number density $N_{\rm e}$ as plasma, the angular frequency of coronal plasma can be written as $\omega_{\rm p} = 5.6 \times 10^{4}\big{(}\frac{N_{\rm e}}{{\rm cm}^{-3}}\big{)}^{1/2}$ ${\rm s}^{-1}$. For a crude approximation when the external magnetic field is not dominant, the refractive index of plasma is given by $n = \big{(} 1 - \omega_{\rm p}^2/\omega^2 \big{)}^{1/2}$, where $\omega$ is the angular frequency of the X-rays in a vacuum. By assuming $r_{\rm cor} = 10r_{\rm g}$, $\tau = 1$--2, and $M=10^{6}$--$10^{8}M_{\odot}$, we can estimate the coronal electron density to be $N_{\rm e} \sim 10^{10}$--$10^{12}$ cm$^{-3}$. This corresponds to the angular frequency of coronal plasma $\omega_{\rm p} \sim 5.6 \times 10^{9}-5.6 \times 10^{10}$ s$^{-1}$. Finally, the deviation of the refractive index of corona from 1 is approximately in the order that $\Delta n \ll 0.5\%$, suggesting the line shifts due to the coronal lensing might be subtle. We caution that this is just a crude approximation since the corona is likely energized by magnetic dissipation, so the true nature of the external magnetic field may play an important role that gives more uncertainty in determining the exact value of $n$. 

Moreover, $n$ can be a complex number consisting of the real and imaginary parts responsible for the light deflection and absorption, respectively. The effects of varying the real part of the refractive index on the motion of light are analogous to varying the curvature of spacetime around black holes. Assuming for simplicity the imaginary part equals 0 means that coronal electrons do not absorb radiation at the particular wavelength of interest. We then note that the effects due to the absorption are still not taken into account here. Instead, this work focuses on the light deflection due to the effects of the coronal plasma mainly induced by the real part of $n$.

A more complex corona such as the one with two thermal Comptonization components \citep[e.g.][]{Petrucci2018} may require the coronal optical properties to vary with its radius and temperature. So does the complex refractive index (e.g., the non-homogenous corona most dense nearest the black hole and thinner further away). This, in turn, may introduce more degeneracy to the model. Furthermore, the deflection of light due to the effects in geometrical optics can be frequency-dependent. This means that the refractive index can vary with the wavelength of light, producing chromatic refraction, which is important if one would like to simulate the full reflection spectrum. We remark that the imaginary part of $n$ responsible for the absorption will determine the flux contribution in each energy band, hence affect the amount of dilution on reverberation features commonly seen in the lag spectra \citep{Emmanoulopoulos2014, Chainakun2016, Epitropakis2016, Wilkins2016, Ingram2019, Caballero2020} as well as in the PSD profiles \citep{Emmanoulopoulos2016, Papadakis2016, Chainakun2019a, Chainakun2021a, Chainakun2021b}. The imaginary part of $n$ that dilutes the flux and produces wavelength-dependent absorption may be important for modelling other timing profiles such as the excess variance which is recently used to probe the intrinsic and absorption variability in AGN \citep{Parker2020, Parker2021}. By treating $n$ as a complex number, it would be possible to model the mean and timing spectra, and simultaneously fit them to the AGN data by, e.g., fine-tuning the real and imaginary parts of $n$.

Finally, the modified metric is not the solution of the vacuum Einstein equation. It gives rise to the energy-momentum tensor of coronal electrons. The calculation for the case of modified Minkowski and Schwarzchild metric by changing $dt \rightarrow n'(r)dt,$ where $n'(r)$ is a smooth cut-off function 
\begin{equation}
n'(r) = \left\{\begin{array}{ll}
                       \frac{1}{\sqrt{\epsilon\mu} }  & ; r< R-\delta \\
                    1- \left(1- \frac{1}{\sqrt{\epsilon\mu} } \right)\sin^2{\frac{\pi(r-R)}{\delta}}    & ; R-\delta\leq r \leq R \\
                        1 & ; r > R
            \end{array} \right.
\end{equation}
shows that $T_{tt}=0$, while $T_{ii}\not= 0$ at $r=r_{\rm cor}$ in both cases. Note that for small $\delta $, $n'(r)$ approaches $n(r)$ in equation \eqref{rf-index}. This corresponds to matter with negligible energy density (comparing to the mass of the black hole), but non zero pressure. We then suspect that our modified Kerr metric would yield a similar result that the coronal electron has negligible energy density but non zero pressure. In principle, using this approach, the nature of corona in terms of its energy and density would naturally depend on the assumption and implementation of $n(r)$ that is used to produce the emission lines. The more realistic values of energy density and pressure associating with a more complex function of $n(r)$ require sophisticated calculations  \citep{deCarvalho2016}, which is a subject of future research.

\section{Conclusion}

The equations for the motions of light rays in the presence of corona with $n \neq 1$ around the accreting black hole are derived. We trace the light rays moving in a modified Kerr metric where the trajectories of the rays are determined by the combined effects of the black hole gravity and light refraction taking place at the coronal boundary. The deflection of light in this approach is comparable to the spacetime expansion or contraction modulated by $n$. A variety of associating line profiles are computed, depending on the inclination angle, the coronal size, the refractive index of the corona, and the disc emissivity index. The lensing corona can induce line shifts (towards higher energy for $n>1$ and lower energy for $n<1$), and thus potentially affect the variability of the emission lines. 

The shapes of the lines should also depend on the distribution function of the hot electrons inside the corona. Chromatic refraction at the coronal boundary can occur due to the geometrical optic effect because the refractive index may depend on the photon frequency. The difficulties are still on the uncertain nature of the corona itself (e.g. what is its optical property and how to robustly measure it). Perhaps, the potential way to search for the effects of the lensing corona on the line profile is to study the AGN with evidence of extended corona observed with high inclinations than $60^{\circ}$. The corona should be radially extended at least $\sim 5r_{\rm g}$ for the lensing effects to be prominent. The concentrated corona smaller than that reduces the chance of the photons to intercept the coronal regime, hence the deviation due to line shifts may not be significant. Our results suggest that this effect on the line shifts would be more drastic ($\Delta g_{\rm max} \gtrsim 0.01$) if the differences in the refractive index of the corona and that of the empty space is $\Delta n \gtrsim 0.5\%$. We remark that, due to the uncertain nature of the X-ray corona, calculating the exact value of its refractive index may not be straightforward. From the crude approximation, the departure of $n$ from unity for the X-ray corona is probably not that large, so these effects may not play a major role in altering the line shifts and time delays.

\section*{Acknowledgements}
We thank the reviewer, Javier A. Garc{\'\i}a, for useful comments and suggestions that help clarify the manuscript. This work was supported by (i) Suranaree University of Technology (SUT), (ii) Thailand Science Research and Innovation (TSRI), and (iii) National Science Research and Innovation Fund (NSRF), project no. 160355. AW acknowledges financial support from National Astronomical Research Institute of Thailand. The numerical calculations were carried out using the BLUECRYSTAL supercomputer of the Advanced Computing Research Centre, University of Bristol. 

\section*{Data availability}
The derived data and models underlying this article will be shared on reasonable request to the corresponding author.

\label{lastpage}

\appendix
\section{Modified Kerr metric}
Consider light travelling a distance $dl$ in a medium with refractive index $n$,  the optical path length (OPL) is defined by $dL:=ndl$. Suppose the light travels this distance in time $dt$ (note that $c=1$), then we define the analogue of the OPL in Minkowski spacetime by
\begin{equation*}
d\tilde{L}^2:=n^2ds^2=n^2\left(-\frac{dt^2}{n^2}+dl^2\right)=-dt^2+(ndl)^2=0.
\end{equation*}
As $n=\sqrt{\varepsilon\mu}$ , the spacetime line element $ds$ can be interpreted as distance in a curved spacetime modified by the medium in Minkowski spacetime. The modified metric can be rewritten as
\begin{equation*}
g_{\mu\nu} = e^a_\mu e^b_\nu \eta_{ab}~,
\end{equation*}
where the transformation $e^a_\mu={\rm diag}\{1/n,1,1,1\}$. Assuming that the distribution of the medium does not affect the background spacetime, thus, we may apply this transformation with any curved spacetime metric   
\begin{equation}
g_{\mu\nu}=e^a_\mu e^b_\nu \bar{g}_{ab}~.
\end{equation}
If we choose $\bar{g}_{ab}$ to be the Kerr metric, one obtains the modified metric presented in eq.~\eqref{Kerr}. Note that, in the case of co-moving fluid in Minkowski, Schwarzshild and FRW spacetime our method yields the comparable metric as, e.g., \cite{Gordon1923} and \cite{Chen2009}. 
\end{document}